\documentclass{iaus}
\usepackage{graphicx}

\title[What triggers star formation in galaxies?] 
{What triggers star formation in galaxies?}

\author[Bruce G. Elmegreen]   
{Bruce G. Elmegreen} \affiliation{IBM Research Division, T.J. Watson
Research Center \\
1101 Kitchawan Road, Yorktown Heights, NY 10598 USA \\ email: {\tt bge@us.ibm.com} \\[\affilskip]
}

\pubyear{2011} 
\volume{284}  
\pagerange{1--12}
\jname{The Spectral Energy Distribution of Galaxies}
\editors{R.J. Tuffs \&  C.C.Popescu, eds.}
\begin{document}

\maketitle

\begin{abstract}
Processes that promote the formation of dense cold clouds in the
interstellar media of galaxies are reviewed. Those that involve
background stellar mass include two-fluid instabilities, spiral density
wave shocking, and bar accretion. Young stellar pressures trigger gas
accumulation on the periphery of cleared cavities, which often take the
form of rings by the time new stars form. Stellar pressures also
trigger star formation in bright-rim structures, directly squeezing the
pre-existing clumps in nearby clouds and clearing out the lower density
gas between them. Observations of these processes are common. How they
fit into the empirical star formation laws, which relate the star
formation rate primarily to the gas density, is unclear. Most likely,
star formation follows directly from the formation of cold dense gas,
whatever the origin of that gas. If the average pressure from the
weight of the gas layer is large enough to produce a high molecular
fraction in the ambient medium, then star formation should follow from
a variety of processes that combine and lose their distinctive origins.
Pressurized triggering might have more influence on the star formation
rate in regions with low average molecular fraction. This implies, for
example, that the arm/interarm ratio of star formation efficiency
should be higher in the outer regions of galaxies than in the main
disks.

\end{abstract}

\firstsection 
\section{The Galactic Scale}

When we observe a star-forming region we sometimes wonder how it got
there, or why stars formed there, or anywhere for that matter. The
average gas density in most galaxies is very low, close to the tidal
limit of $\rho_{\rm tidal}=-1.5\Omega R (d\Omega/dR)/(\pi G)\sim1$
cm$^{-3}$ for galactic rotation rate $\Omega$ and radius $R$. This is
also about the gas surface density limit where Toomre $Q=\sigma_{\rm
gas}\kappa/(\pi G\Sigma_{\rm gas})\sim1$ for velocity dispersion
$\sigma_{\rm gas}$, epicyclic frequency $\kappa$, and mass column
density $\Sigma_{\rm gas}$. Stellar explosions produce denser gas in
supernova remnants, but this gas is usually too tenuous to be strongly
self-gravitating, as in the Veil Nebula, and it is also too short-lived
in that state for self-gravity to operate (\cite[Desai et al.
2010]{desai10}). The two-phase instability makes cool gas, but this
operates only between temperatures of $\sim10,000$K and $\sim100$K,
which is not cold enough to make stars. Star formation requires a
thermal Jeans mass close to a solar mass. The thermal Jeans mass is
$M_{\rm J,th}= \rho k_{\rm J}^{-3}\sim7(T/10\;{\rm K})^{1.5}n^{-0.5}
M_\odot$ for thermal Jeans wavenumber $k_{\rm J}=(4\pi G\rho)^{1/2}/c$
and isothermal sound speed $c=(kT/\mu)^{1/2}$ with mean weight $\mu$ of
atomic gas, and density $n=\rho/\mu$. For the range of temperatures
expected from the thermal instability and for the average density,
$M_{\rm J,th}\sim 200M_\odot$, which is too large for a star. Star
formation requires gas that is both cold and dense to bring $M_{\rm
J,th}$ down. Processes that do this may be thought of as star-formation
triggers.

A fair question is whether the ISM can make cold and dense gas on its
own, without non-ISM processes, such as stellar density waves, galactic
shear, superbubbles, etc.. If there were no stars to compress the gas
or supplement its gravitational self-attraction with additional mass,
would new stars form? The answer is probably yes because galaxies once
had no stars, and the gas formed stars anyway. Still, galaxy collisions
in the young universe could have triggered the star formation by
forcing the gas to be dense in the shocked overlap region and in a
central concentration that formed after angular momentum loss.

During the last few years, observations of very young galaxies seem to
suggest that they can form giant clumps of star formation on their own,
even with no observable underlying disk of older stars and no evidence
for an interaction with another galaxy (\cite[Elmegreen et al.
2009ab]{elmegreen09a}, \cite[Genzel et al. 2011]{genzel11}). The only
possible process for this seems to be a gravitational instability in
the whole gas disk, with a resulting clump mass comparable to the
turbulent Jeans mass, $M_{\rm J,turb}=\Sigma_{\rm gas}k_{\rm
J,2D}^{-2}$ for $k_{\rm J,2D}=\pi G \Sigma_{\rm gas}/\sigma_{\rm
gas}^2$. The clumps in young galaxies are large relative to the galaxy
radii because the turbulent speeds are large relative to the rotation
speeds $v_{\rm rot}$ (\cite[Davies et al. 2011]{davies11}). This
follows from $Rk_{\rm J,2D}\sim GM/(R\sigma_{\rm gas}^2)\sim (v_{\rm
rot}/\sigma_{\rm gas})^2$.

This process of gravitational collapse of the ISM is the most
fundamental of triggering mechanisms. To carry the initial collapse all
the way to stars in a young ISM may have also relied on the pervasive
presence of CO molecules (\cite[Daddi et al. 2010]{daddi10},
\cite[Tacconi et al. 2010]{tacc10}), which allows for cooling to
$T<100$K, and on the extremely high $\Sigma_{\rm gas}$, which lowers
$M_{\rm J,th}$ by increasing the average midplane density. If we write
the midplane pressure in a pure-gas disk as $P=0.5\pi G\Sigma_{\rm
gas}^2$ and the thermal Jeans mass as $M_{\rm J,th}=c^4(4\pi
G)^{-1.5}P^{-0.5}$, then
\begin{equation}
M_{\rm J,th}=3.8(T/100\;{\rm K})^{2}(\Sigma_{\rm gas}/100M_\odot\;{\rm
pc}^{-2})^{-1}M_\odot\label{jeans}\end{equation} from which it can be
seen that the observed ISM column densities in the clumps of young
galaxies, $\Sigma_{\rm gas}> 100M_\odot$pc$^{-2}$ (\cite[Tacconi et al.
2010]{tacc10}), raise the pressure so much that even temperatures of
100 K may be low enough for star formation. Collapse inside these
clouds raises the pressure more because of the higher local column
density in the resulting core, and it lowers the temperature more
because of opacity.

When stars are present, gravitational instability in the ISM is
augmented by stellar gravity. This happens most commonly in three ways:
by a two-fluid swing-amplified instability (\cite[Toomre
1981]{toomre81}), in the shocks of global spiral density waves
(\cite[Roberts 1969]{roberts69}), and in the dense central regions
formed by gas accretion in a bar potential (\cite[Matsuda \& Nelson
1977]{matsuda77}).

The first of these involves the simultaneous collapse of gas and stars,
which produces a moderately dense clump of both components on a
timescale $(G\rho_{\rm total})^{-1/2}\sim\kappa^{-1}$ for $Q\sim1$ .
The stellar part of this collapse bounces as the stars move away with
enhanced energy in epicycles. Nearby stars form a spiral wake
(\cite[Julian \& Toomre 1966]{julian66}).  The dissipative gas remains
as a dense clump at the center of gravity of the instability, with a
gas filament also forming in the wake. This is the two-fluid
swing-amplified instability (\cite[Jog \& Solomon 1984]{jog84},
\cite[Rafikov 2001]{rafikov01}, \cite[Romeo \& Wiegert 2011]{romeo11},
\cite[Elmegreen 2011]{e11}). It appears to be common in galaxies,
taking the form of multiple spiral arms that are long and irregular in
the outer parts. Usually these arms are more regular in the inner
regions, where multiple-arms often merge into symmetric two-arm
structures midway in the disk (\cite[Elmegreen \& Elmegreen
1995]{e95}).

\begin{figure}[b]
\begin{center}
 \includegraphics[width=5.in]{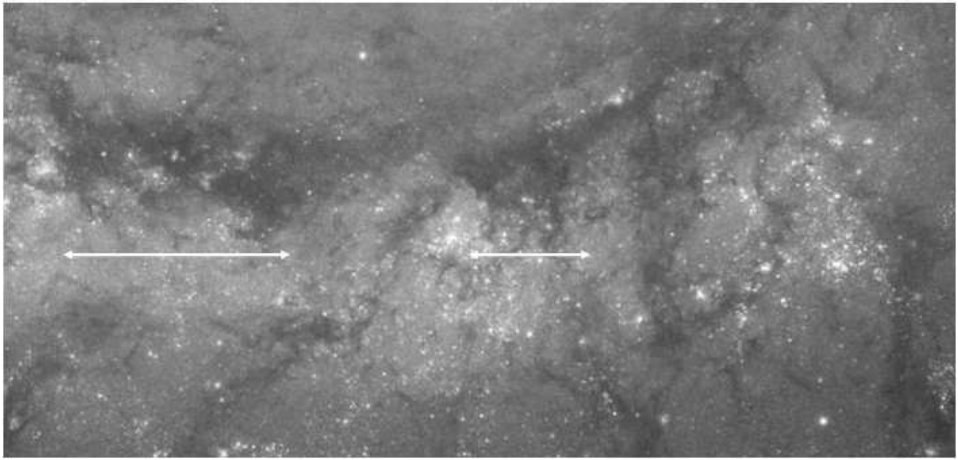}
\caption{HST image of the southern arm in M51, showing large dark
clouds, star formation in the clouds, and young stars downstream with
shells and rings around them. The gas flows from the top left to the
bottom right in the figure. The large dark clouds contain around
$10^7M_\odot$ of gas. The shells and rings appear to contain new star
formation along the periphery.}
   \label{m51_fullresolution_southinnerarm_withlines_BW}
\end{center}
\end{figure}

The second way in which stellar gravity augments the condensation of
gas into star-forming clumps is through stellar density waves, which
arise in the gas+star mixture as a result of perturbations from the
outer disk, companion galaxies, or bars, and which may last for several
rotations (\cite[Lin \& Shu 1964]{lin64}) before moving to the center
(\cite[Toomre 1969]{toomre69}). These waves move through the gas and
stars (unlike the swing amplifier instability which follows the
gas+star mixture for a time $\kappa^{-1}$) and they shock the gas as it
passes through. The shock appears as a thin dust lane. If the shocked
gas is dense enough, it can become gravitationally unstable and form
giant clumps and star complexes. It can also form stars by cooling
interarm gas and squeezing interarm clouds.  Most star formation in the
Milky Way is in giant molecular clouds (GMCs) that are parts of
$10^7M_\odot$ HI+CO complexes in spiral arms (\cite[Grabelsky et al.
1987]{grabelsky87}). Because shear is low in spiral density wave arms,
the process of gas collapse can be augmented by magnetic tensional
forces, which remove angular momentum (\cite[Elmegreen 1987]{e87},
\cite[Kim et al. 2002]{kim02}).

Some galaxies have a stellar component that is either too warm or too
rapidly rotating for its column density to form strong spiral arms,
i.e., $Q_{\rm stars}>>1$; then the gas+star mixture is not particularly
unstable. The gas is dissipative, however, and can still collapse on
its own, or with only a small contribution from underlying stars.  The
result is a network of thin and short gaseous arms that form stars,
giving a flocculent appearance. These arms should be co-moving with the
material close to the site of the initial collapse, although they can
be wave-like far from this site, in the spiral wake.

\begin{figure}[b]
\begin{center}
 \includegraphics[width=5.in]{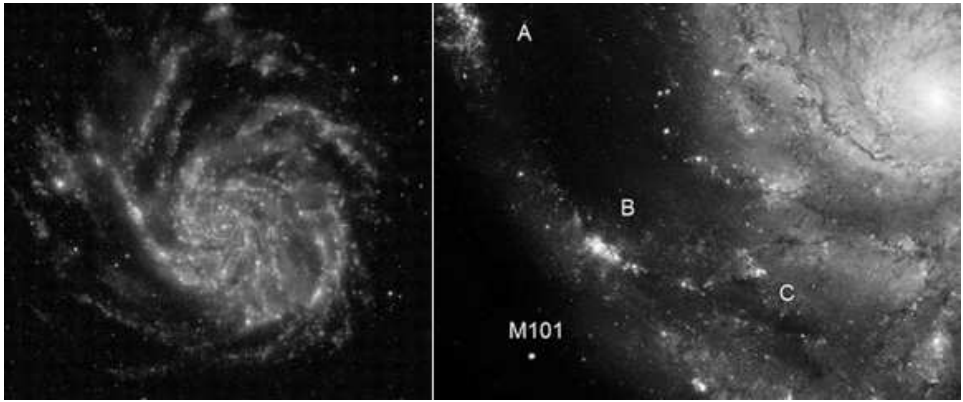}
\caption{(Left) GALEX image in the uv of M101 from \cite[Gil de Paz et
al. (2007)]{gil07}.  (Right) HST image of the western part of M101,
showing two giant star complexes (A and B) and a dark cloud without
much star formation yet (C). Young stars are centered in the spiral
arms of M101, suggesting these arms are material patterns made from
local gravitational collapse. }
   \label{m101_composite_BW}
\end{center}
\end{figure}

Figure \ref{m51_fullresolution_southinnerarm_withlines_BW} shows a
piece of the galaxy M51 from an image taken with the Hubble Space
Telescope. There is a spiral density wave with two large concentrations
of gas in the dust lane, $10^7M_\odot$ each, and star formation in each
concentration. When there are several of these concentrations along
part of an arm, they look like ``beads on a string''. \cite[Efremov
(2010)]{efremov10} studied the relation between such regular patterns
and the magnetic field.  For several magnitudes of extinction, and for
a thickness through the plane of $\sim100$ pc, the average density in
the dustlane is $\sim10$ cm$^{-3}$ (\cite[D. Elmegreen 1980]{e80}) and
so the average dynamical time is $\sim(G\rho)^{-1/2}\sim30$ Myr. This
average density is about what is expected for a density wave shock
where the average incident density is the intercloud value ($\sim0.3$
cm$^{-3}$) and the ratio of the shock velocity to the internal
turbulent speed is $\sim5$. The average dynamical time is too long for
gravitational collapse to develop much while the gas is in the dust
lane, but the interarm gas is clumpy and these clumps feel a pressure
increase when they enter the arm, causing them to collapse. \cite[Kim
\& Ostriker (2002)]{kim02a} showed the process of dustlane collapse
starting with a smooth ISM. It took several rotations to build up
enough density structure for a self-gravitating cloud to form in a
spiral arm.

\begin{figure}[b]
\begin{center}
 \includegraphics[width=4.in]{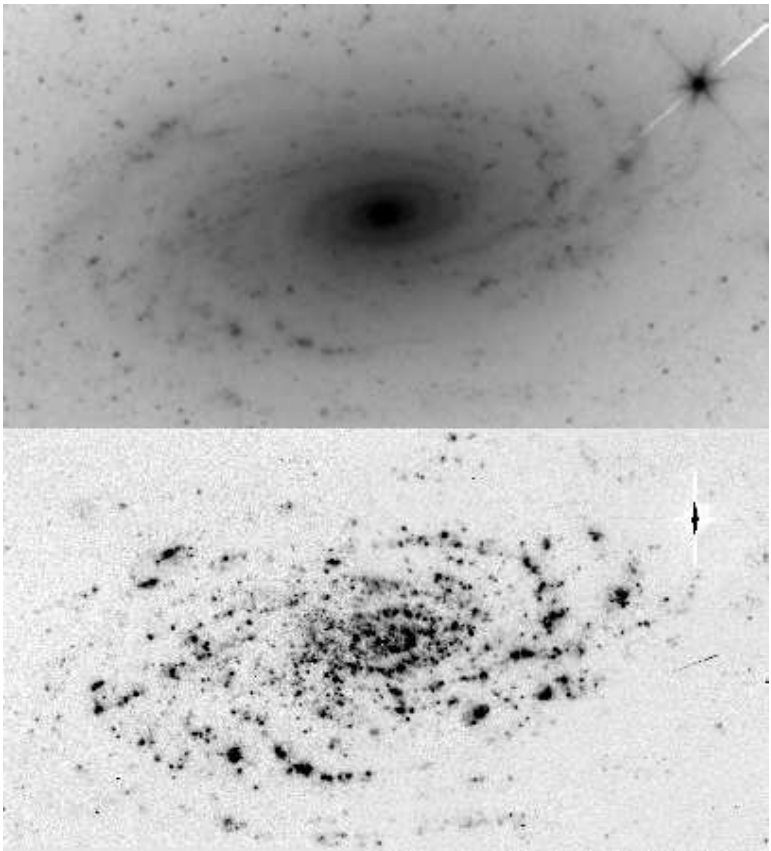}
\caption{(top) Spitzer $3.6\mu$ image from \cite[Elmegreen et al.
(2011)]{elm11} showing weak and irregular underlying stellar waves.
(bottom) H$\alpha$ image from \cite[Dale et al. (2009)]{dale09} showing
numerous beads on a string of star formation.}
   \label{n5055_composite_BW}
\end{center}
\end{figure}

The structure of gas in the interarms is a remnant of the structure it
had leaving the previous arm, which can also be seen in Figure 1. As
the ISM moves through the arm, star formation pressures from HII
regions, winds, and multiple supernovae make superbubbles and
super-rings. The figure shows many rings of dust. The small rings can
be three-dimensional shells seen with enhanced absorption along the
lines of sight through the edges, but the large rings are much bigger
than the ISM scale height and have to be two-dimensional. All of these
structures seem to contain active star formation in the peripheral
dense gas, in addition to OB associations and star complexes inside the
cleared regions. Some of the active star formation is {\it lingering}
in the dense clouds, which means that it follows the dense gas as it
moves. Other active star formation on the periphery is probably {\it
triggered} by the high pressures that made the cavities. This is a
second type of triggering for star formation: sequential triggering by
previous generations of stars. We review this in the next section.

Figures \ref{m101_composite_BW} and \ref{n5055_composite_BW} show a
multiple-arm galaxy, M101, and a flocculent galaxy, NGC 5055. The
spiral structure in M101 is more irregular than in the grand design
galaxy M51, but there are still long spiral arms and each arm has
several regularly-spaced giant star complexes in it, in various stages
of collapse. The right-hand side of Figure 2 highlights the western arm
in M101, suggesting that the giant dust cloud toward the bottom is in a
younger stage than the two other star complexes toward the west. Unlike
in M51, the star formation in M101 seems to be centered on the arms
with no rings offset to one side. This is expected when the whole arm
is the result of a gravitational instability, because it all twists
around with local shear and has little relative motion between the
pattern speed and the gas.  Figure 3 shows numerous long and thin
spiral arms with more of the "beads on a string" pattern. These arms
are so thin and irregular that they should be mostly gas. They shear
around, stretch out, and disperse over time, most likely forming stars
by gravitational collapse along their length.  Rings and shells are not
visible in this image. It would be interesting to observe further
whether flocculent galaxies produce the same type of ring pattern as
observed down stream from the spiral arms in M51.

\begin{figure}[b]
\begin{center}
 \includegraphics[width=4.in]{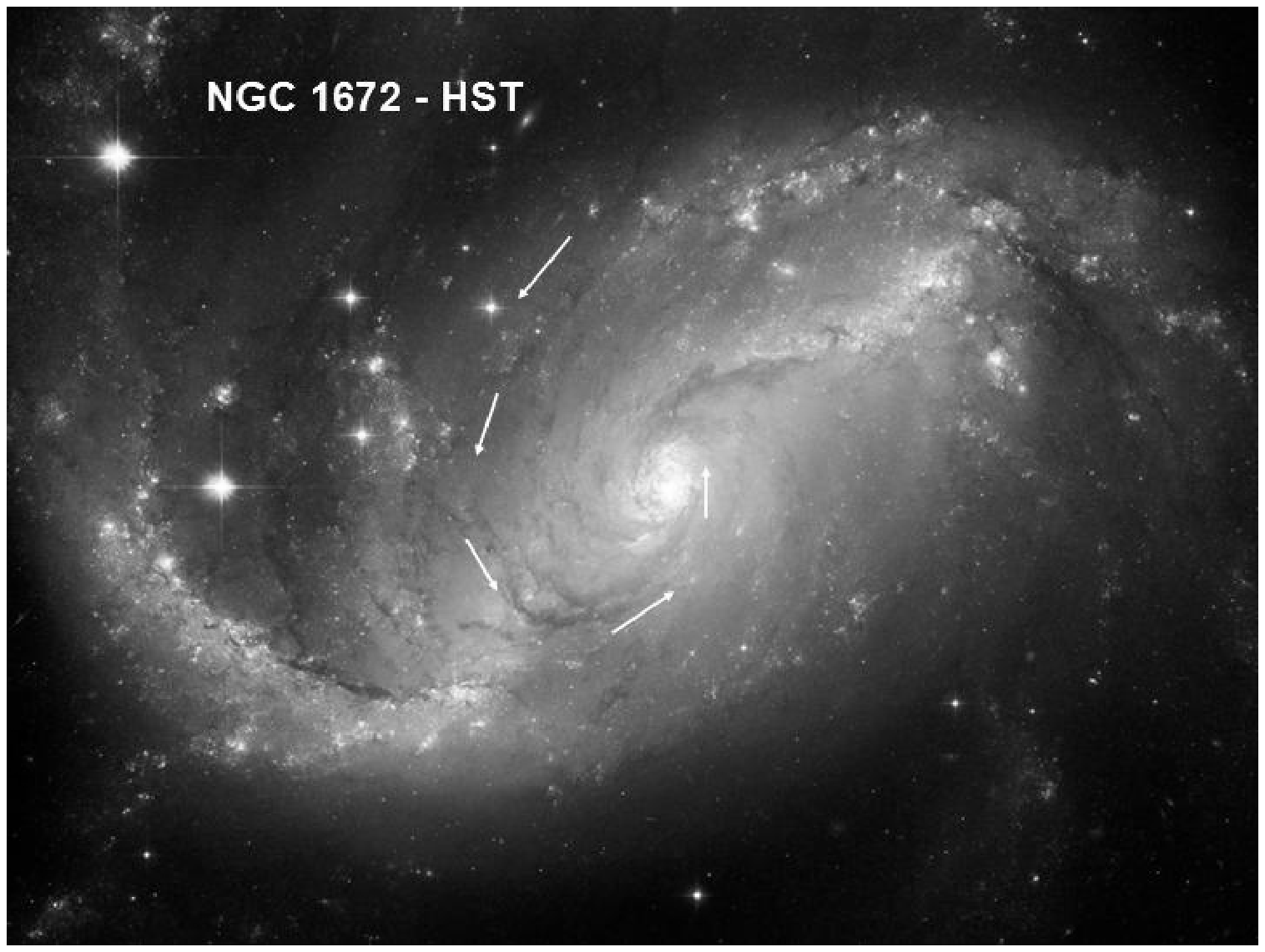}
\caption{Barred galaxy NGC 1672 from the Hubble Heritage Team. Dust
lanes suggest the gas flow pattern toward the center (arrows), where
star formation is very active.}
   \label{preston11f_n1672}
\end{center}
\end{figure}

The third common way in which stars in a galaxy promote gravitational
collapse in the gas is through bar-driven inflows. Bars exert strong
negative torques inside their corotation radii, which are typically at
1.2 to 1.4 bar radii. Inside this, the gas hits the bar from behind and
shocks, forming a dustlane (\cite[Athanassoula 1992]{athan92}). The
shock strips angular momentum and orbital energy from the gas, which
then falls along the bar to the region of the inner Lindblad resonance
(ILR). After sufficient accumulation, the gas becomes unstable and form
stars. Often it does this in a ring or tight two-arm spiral. There may
be a second, smaller bar inside the ILR which affects the orbits there
and brings the gas in further.

Figure \ref{preston11f_n1672} shows how the gas flows to the center of
NGC 1672.  Dust streamers plunge from near the end of the bar into the
bar itself, and then directly to the center. This direct path means
that gas inflow is rapid. A comparison of the inflow rate to the gas
reservoir gives the timescale for accretion. In a study of NGC 1365
(\cite[Elmegreen et al. 2009c]{e09c}), which looks like NGC 1672, the
timescale for gas exhaustion is only $\sim0.5$ Gyr. If the initial
reservoir of gas was comparable to what is there today, then this short
time is also comparable to the age of the bar. That is, bars like these
have to be fairly young, $\sim1$ Gyr, or the gas would have been
cleared out by now.

NGC 1365 and many other barred galaxies have massive clusters in their
ILR regions -- much more massive than typical disk clusters or clusters
forming in spiral density wave arms. Perhaps this difference is only a
size-of-sample effect if the ILR regions of bars have higher overall
star formation rates. Or, star formation in bar ILRs may differ from
that in main disks.

\section{Super-Bubbles, Super-Rings, and Colliding Flows}

Figure \ref{m51_fullresolution_southinnerarm_withlines_BW} shows a
spiral arm segment with large rings of gas and star formation
downstream from the arms. These are giant cavities are made by star
formation, as discussed above. The solar neighborhood is downstream
from the Carina spiral arm and also has a large cavity apparently made
by star formation. This is associated with the expanding HI ring
discovered by \cite[Lindblad et al. (1973)]{lind73} with a hole in the
diffuse dust (\cite[Lallement et al. 2003]{lallement03}). The size of
the dust hole is 100 pc by 300 pc, and it extends all the way through
the disk. The cavities in Figure
\ref{m51_fullresolution_southinnerarm_withlines_BW} are about the same
size, and larger than the ISM thickness.  Thus they are rings rather
than shells. The smaller regions could be three-dimensional shells.

Expanding rings are more unstable than expanding shells because the
self-gravitational force vectors that drive gas collection in a
one-dimensional section of a ring are all directed toward the growing
condensation.  According to \cite[Elmegreen (1994)]{e94}, the time for
significant collapse in an expanding shell is $\sim100(n{\cal
M})^{-0.5}$ Myr for ambient density $n$ in cm$^{-3}$ and Mach number
${\cal M}$, and the time in an expanding ring is $124(n{\cal
M}^2)^{-0.5}$ Myr. The ring expression contains a stronger dependence
on the Mach number, and this give the ring a shorter collapse time.

Generally, the expansion scale equals the shock speed multiplied by the
time, and the shock speed is about $(P/\rho_0)^{0.5}$ for driving
pressure $P$ and preshock density $\rho_0$. If the relevant time is the
collapse time, $(G\rho_{\rm comp})^{-0.5}$ for compressed density
$\rho_{\rm comp}$, then the expansion scale is, after rearrangement,
$(P/\rho_{\rm comp})^{0.5}(G\rho_0)^{-0.5}$, which is the velocity
dispersion in the compressed region multiplied by the dynamical time in
the ambient gas.  Details about the compression source drop out.

If the expansion scale is less than the cloud scale, then pillars and
bright rims form by the push-back of interclump gas (\cite[Elmegreen,
Kimura \& Tosa 1995]{e95a}, \cite[Gritschneder et al. 2009]{grit09}).
Star formation is a fast process (squeezing pre-existing clumps), the
velocity of triggered stars is small, and causality is difficult to
determine, i.e., stars could have formed in the clumps anyway.  If the
expansion scale is larger than the cloud scale, then shells and rings
form by the push-back of all ISM gas (e.g. \cite[Dale et al.
2011]{dale11}), triggering is a slow process because new clumps have to
form on a timescale of $(G\rho_{\rm shell})^{-0.5}$, and the velocity
of triggered stars can be large, on the order of the shock speed,
$(P/\rho_0)^{1/2}$.  There is also a clear causality condition because
two stellar generations have to be separated by a distance equal to the
age times the mean velocity.

A popular cloud formation scenario considers the compression of gas
between two ``colliding flows'' (e.g., \cite[Heitsch et al.
2008]{heitsch08}, \cite[Audit \& Hennebelle 2010]{audit10}). Shell or
ring accumulation forms clouds too, but is different in several ways.
Shells and rings have a lateral expansion as they expand radially, and
this lateral expansion resists gravitational collapse. Shells and rings
also decelerate so that newly formed condensations protrude out the
front and have the potential to erode. Shells and rings have a shock on
only one side. A shock boundary condition causes a diverging flow at
each clump, and this divergence resists collapse. The pressure boundary
condition on the other side of the shell or ring squeezes the
perturbations and aids collapse. For colliding flows with constant
velocity streams, there is a shock on each side and no acceleration of
the condensation between them when it is in equilibrium. We observe
shells and rings commonly, as shown in the figures and in surveys
(e.g., \cite[K\"onyves et al. 2007]{konyves07}, \cite[Ehlerov\'a \&
Palou\v s 2005]{ehlerova05}), but there is no clear evidence yet for
cloud formation on GMC-scales by colliding flows (on much larger
scales, the collision between two galaxies can have a colliding flow;
\cite[Herrara et al. 2011]{herrara11}). Still, colliding flows are a
good model to study molecule formation and fragmentation in a dynamic
environment.

Triggering in the ring RCW 79 was studied in detail by \cite[Zavagno et
al. (2006)]{zavagno06}, who suggested there was a collapsed neutral
region containing young stars, 0.1 Myr old, along the periphery of a
swept-up region that was 1.7 Myr old. There are several neutral
condensations in this shell, somewhat equally spaced around part of the
projected edge.  \cite[Deharveng et al. (2010)]{deharveng10} studied
102 Milky Way bubbles and suggested that 18 of them have either
ultracompact HII regions or methanol masers along their edge,
suggesting triggering of massive stars in swept-up gas.

\section{Bright Rims and Pillars}

\begin{figure}[b]
\begin{center}
 \includegraphics[width=4.in]{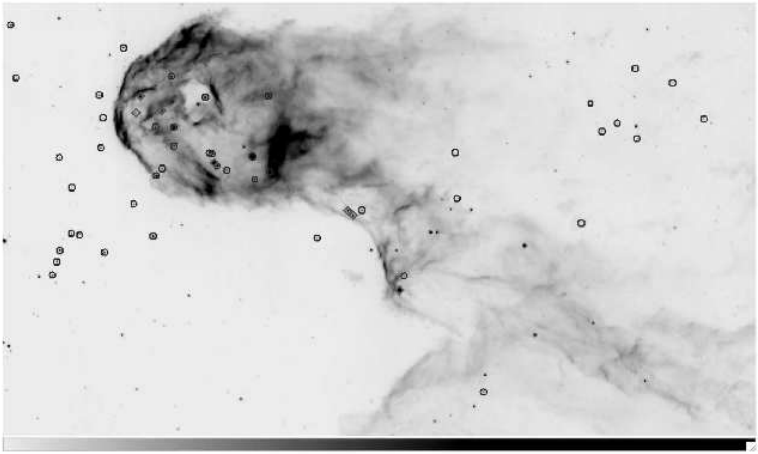}
\caption{Bright rim at the edge of the HII region IC 1396, from
\cite[Reach et al. (2009)]{reach09}. The youngest stars are denoted by
diamonds. They were probably triggered to form in compressed cloud
clumps that were exposed by ionization and movement of the lowest
density, surrounding regions.} \label{reach2009f16}
\end{center}
\end{figure}

Most regions of massive star formation have bright ionized rims and
pillars of neutral gas pointing to the sources of radiation. A
well-studied example is IC 1396, which is a circular HII region 12 pc
in radius, with an expansion speed of 5 km s$^{-1}$ and an age of 2.5
Myr (\cite[Patel et al. 1995]{patel95}). On the western edge is a
neutral pillar several pc long (Fig. \ref{reach2009f16}) with very
young (class I -- diamonds) stars in head-like protrusions and other
young stars (class II -- circles) all around (\cite[Reach et al.
2009]{reach09}).  This is a typical case.

There is often an age gradient in pillars with the oldest stars closer
to the center of the HII region (\cite[Sugitani et al.
1995]{sugitani95}).  The timescale for triggering can be very fast,
$10^4$ yrs (\cite[Sugitani et al. 1989]{sugitani89}), but the age
gradient can span a time of $10^6$ years or longer during which the
compression moves down the pillar (\cite[Smith 2010]{smith10}).

This mechanism of triggering was originally illustrated in models by
\cite[Klein et al. (1983)]{klein83}. Recent simulations by \cite[Bisbas
et al. (2011)]{bisbas11} compare radiative implosion in a variety of
conditions. At low incident flux, the implosion is slow, the bright rim
is wide, and star formation is far from the tip in the center of a
converging compression front. At high incident flux, the implosion is
fast, the pillar is narrow, and star formation is close to the tip
where the initial compression was concentrated. At late times, the
pillar can be long and thin, with bare stars near the head and other
stars embedded throughout. If there are many small clumps in the
original cloud, then there can be many small pillars, one for each
clump, with star formation in each.

\section{The Antenna Galaxy: a Mixture of Triggering Processes}

\begin{figure}[b]
\begin{center}
 \includegraphics[width=4.in]{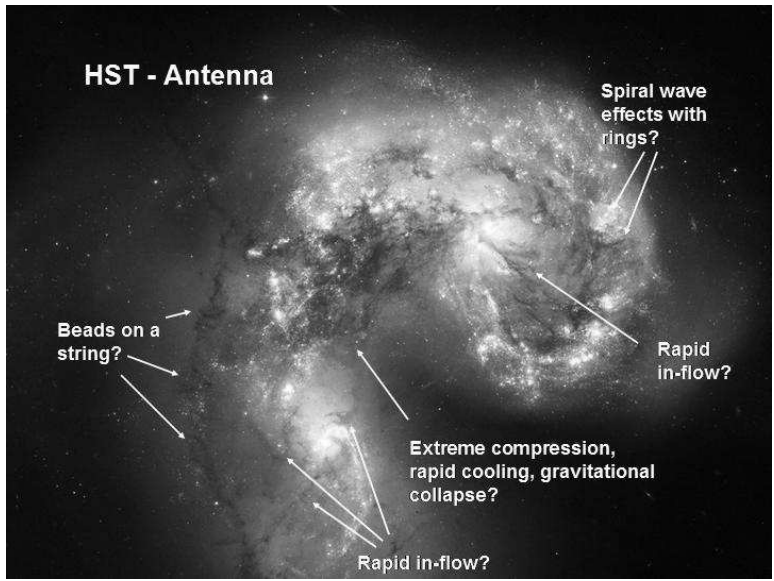}
 \caption{HST image of the Antenna galaxy showing several types of
triggering discussed in the text.} \label{hubble_antenna}
\end{center}
\end{figure}

Several of the triggering mechanisms discussed above are illustrated in
the Antenna galaxy, which is a merger. Figure \ref{hubble_antenna}
shows the Antenna labeled with suggested analogies: spiral arm shock
triggering with downstream rings, beads on a string from gravitational
collapse in a filament, radial inflow in elongated, bar-like regions,
and a converging flow that produces extreme compressions between the
two gas disks.

\section{Triggering and Empirical Star Formation Laws}

Star formation triggering is about when, where, and how star formation
begins. The global laws of star formation, such as the \cite[Kennicutt
(1998)]{ken98} or \cite[Bigiel et al. (2008)]{bigiel08} relations, show
a correlation between the star formation rate and the supply of gas of
various types. The \cite[Kennicutt (1998)]{ken98} study shows a star
formation rate per unit area that scales with the 1.4 power of the
total gas column density in the main disks and central regions of
spiral galaxies, and the \cite[Bigiel et al. (2008)]{bigiel08} and
\cite[Leroy et al. (2008)]{leroy08} studies show a linear relation
between the star formation rate per unit area and the CO emission per
unit area. A linear relation for dense gas traced by HCN was first
shown by \cite[Gao \& Solomon (2004)]{gao04} and \cite[Wu et al.
(2005)]{wu05}.

In all cases, star formation is assumed to occur only in cold dark gas,
which is traced by CO at low density and HCN and other molecules at
high density. The empirical laws suggest that the star formation rate
is independent of how the gas is assembled, and therefore independent
of the triggering mechanisms.  This is particularly true if the average
pressure and radiation field determine the general molecular state of
the gas, and the pressures which make shells and rings do not change
this state much, they just push the gas around.  Triggering will also
not influence the empirical laws much if the various types of
triggering are always present to the same degree. For example, if every
initial incidence of star formation is followed by a certain proportion
of additional stars that form by radiative implosion or in shell and
ring collapse, then the conversion efficiency from ambient gas to young
stars will contain all of these effects combined. Each triggering
process will not stand out separately in the total star formation rate.

According to \cite[Gnedin \& Kravtsov (2011)]{gnedin11} and others, CO
and star formation are predictable from the average pressure,
self-shielding, radiation field, and other quantities.  The pressure
that determines molecular self-shielding does not vary much on the
scale of shells and rings once they are old enough for star formation
to begin. Regions with radiative implosion have high pressure, but this
may be a relatively minor star formation event. The pressure varies in
spiral arms, but this should only put scatter in the empirical
relations. Thus the signatures of localized triggering may be
imperceptible in the empirical star formation laws.

\section{All you need is cold gas: the legacy of K.E. Edgeworth}

The universality of the empirical laws combined with the evidence for
star formation in shells, spiral arms, pillars, and other pressurized
structures suggests that the primary ingredient for star formation is
cold gas and not a particular geometry. Cold gas means colder than
thermal equilibrium commonly gives for atomic gas in the background
galactic radiation field. Cold gas for star formation implies shielding
from starlight and the presence of molecules for rotational cooling,
which can operate down to a few degrees above the microwave background.
Once the gas is shielded from starlight and turns molecular, background
pressure and turbulent motion will compress it to a high density in
small regions. Then gravity becomes important and the thermal Jeans
mass drops to be comparable to the region mass. Star formation usually
follows.

Shell, ring, or pillar-like triggering and spiral waves might be more
important as a net positive contributor to star formation in regions
where the molecular fraction is moderately low on average, i.e., at the
boundary between highly molecular inner disks of galaxies and highly
atomic outer disks. In the inner disks, the gas pretty much stays
molecular everywhere because of the high pressure from its weight in
the disk, and because of the high opacity. Starlight ionizes and
photodissociates molecules at the cloud edges (\cite[Heiner et al.
2011]{heiner11}) and between the clouds.  Cold gas is present in most
clouds, so additional pressures from young stars or spiral waves will
not make the gas much colder or more molecular. In the far-outer
regions, however, it is rather difficult to turn atomic gas into
molecular gas because the average pressure and opacity are very low.
Then superbubbles might not trigger new star formation, but only make
diffuse atomic shells which have little cold gas. Between these two
zones, or perhaps in dwarf irregular galaxies where the same
intermediate conditions apply, triggering by the pressures of young
stars might make more of a difference in the total star formation rate.

An observational test of this prediction would be to measure the radial
variation of the ratio of the normalized star formation efficiency in
compressed and non-compressed regions. The normalized efficiency is the
star formation rate per unit gas mass per unit dynamical rate.  An
example would be the ratio of the efficiency in the spiral arms to the
interarm regions. If this ratio increases with radius, then spiral arm
triggering is doing more to enhance star formation in the outer parts
of galaxies, where the gas is mostly atomic, than in the inner parts,
where the gas is mostly molecular. Alternatively, one could plot this
ratio versus the average molecular fraction instead of the radius.

The overriding importance of cold gas for star formation was originally
recognized by Kenneth Essex Edgeworth in 1946 -- long before the modern
star formation era. (For a biography of Edgeworth, see \cite[Hollis
1996]{hollis96}.)  At that time, most stars, like the Earth and the
whole universe, were thought to be about 3 Gyr old, star formation was
not considered to happen in the present day, except possibly for the
most massive stars (which were known to burn their fuel quickly), and
the stability of galactic disks was not yet understood. Edgeworth and
others were thinking about star formation mostly in the context of the
beginning of the universe. He noted from a thermal Jeans analysis that
star formation by gravitational instability in a galaxy disk, using an
equation like eq. \ref{jeans}, requires a temperature of $\sim2.8$K,
which he considered ``very improbable.'' He also said that angular
momentum from galactic rotation was too large for the solar system to
form, and suggested that stars might form instead by successive
condensations to overcome the angular momentum problem. This led him to
``expect to find that the majority of the stars were members of star
clusters'' which is ``not in agreement with observation.''  Finally, he
suggested that the rotating gas disk of a galaxy, at $\sim1000$ K,
breaks up into azimuthal filaments, which then break up into stars
following the removal of heat. He went on to suggest that the residual
material around each star makes planets.

These ideas are all essentially correct by modern standards -- even
though Edgeworth did not believe or observe them at the time. Edgeworth
understood that star formation requires very cold gas, it most likely
occurs in massive aggregates, and it should be patterned as beads on a
string for the galactic scale.  These ideas were too far ahead of their
time to have much influence. Star formation as we know it was
discovered several years later when \cite[Ambartsumian (1949)]{am49}
showed that local OB associations are expanding away from a common
center. This limited their age to 10 Myr.

\section{Summary}

A variety of processes cause interstellar gas to become cold enough and
dense enough to form stars. On galactic scales, stellar instabilities,
spiral waves, and global perturbations like bars can move the gas
around supersonically and cause shocks to form that are larger than the
characteristic size of a gravitational instability in the gas. Then
giant cloud complexes form from the ambient gas. As these complexes
dissipate their internal turbulent energy, they contract
gravitationally and fragment because of converging and diverging
turbulent motions until dense, thermally-dominated cold cloud cores
form. Stellar pressures also compress the gas supersonically. On
sufficiently large scales, these compressions lead to collapse in
shells and rings. On small scales, stellar pressures can turn in
pre-existing clumps unstable to collapse, especially along the edges of
HII regions and super-bubbles.

The empirical laws of star formation have no obvious connection to the
details of these triggering mechanisms. The empirical laws state mostly
that star formation requires cold and dense gas. In the case of the
Kennicutt (1998) law, with its non-linear dependence of star formation
rate on total gas density, empirical evidence suggests also that
general dynamical processes in the ISM are involved in determining the
time scale.  Since triggering on the length scale of these laws has the
same dynamical time scale, all of the various triggering processes can
mix together without much distinction.

The distinct contribution that triggering makes to the star formation
rate might be most evident in large-scale regions where the average
molecular fraction is neither very high nor very low. There the
dynamical processes related to cloud formation could have a significant
influence on the abundance of cold gas in clouds.  There might still be
a linear relation between cold gas mass and star formation rate at
these places, because star formation follows cold gas no matter what
forms the cold gas, but the rate of both cold gas formation and star
formation could be modulated by dynamical processes more there than
elsewhere.

\vspace{0.5cm}

The author is grateful to the National Science Foundation for support
from grant AST-0707426, and to the conference organizers, particularly
Professors Cristina Popescu and Richard Tuffs, for their support and
hospitality.


\begin{thebibliography}{}

\bibitem[Ambartsumian (1949)]{am49} Ambartsumian, V.A. 1949, \textit{Soviet
\textit{AJ}}, 26, 3

\bibitem[Athanassoula (1992)]{athan92} Athanassoula, E. 1992,
\textit{MNRAS}, 259, 345

\bibitem[Audit \& Hennebelle (2010)]{audit10} Audit, E., \&
Hennebelle, P. 2010, \textit{A\&A}, 511, 76

\bibitem[Bigiel et al.(2008)]{bigiel08} Bigiel, F., Leroy, A., Walter,
F., et al. 2008, \textit{AJ}, 136, 2846

\bibitem[Bisbas et al.(2011)]{bisbas11} Bisbas, T.G., Wünsch, R.,
Whitworth, A.P., Hubber, D.A., Walch, S. 2011, \textit{ApJ}, 736, 142

\bibitem[Daddi et al. (2010)]{daddi10} Daddi, E. et al. 2010a,
\textit{ApJ}, 713, 686

\bibitem[Dale et al. (2009)]{dale09} Dale, D.A. et al. 2009,
\textit{ApJ}, 703, 517

\bibitem[Dale et al. (2011)]{dale11} Dale, J.E., W\"unsch, R., Smith,
R.J., Whitworth, A., \& Palou\v s, J. 2011, \textit{MNRAS}, 411, 2230

\bibitem[Davies et al. (2011)]{davies11} Davies, R., F\"orster
Schreiber, N.M., Cresci, G., et al. 2011, \textit{ApJ}, 741, 69

\bibitem[Deharveng et al.(2010)]{deharveng10} Deharveng, L., Schuller,
F., Anderson, L. D., et al. 2010, \textit{A\&A}, 523, 6

\bibitem[Desai et al. (2010)]{desai10} Desai, K.M., Chu, Y.-H., et al.
2010, \textit{AJ}, 140, 584

\bibitem[Edgeworth (1946)]{edgeworth46} Edgeworth, K.E. 1946,
\textit{MNRAS}, 106, 470

\bibitem[Efremov (2010)]{efremov10} Efremov, Yu. N. 2010,
\textit{MNRAS}, 405, 1531

\bibitem[Ehlerov\'a \& Palou\v s, (2005)]{ehlerova05} Ehlerov\'a, S.,
\& Palou\v s, J. 2005, \textit{A\&A}, 437, 101

\bibitem[Elmegreen (1987)]{e87} Elmegreen, B.G. 1987, \textit{ApJ},
312, 626

\bibitem[Elmegreen (1994)]{e94} Elmegreen, B.G. 1994, \textit{ApJ},
427, 384

\bibitem[Elmegreen (2011)]{e11} Elmegreen, B.C. 2011, \textit{ApJ},
737, 10

\bibitem[Elmegreen, Kimura \& Tosa (1995)]{e95a} Elmegreen, B.G.,
Kimura, T., \& Tosa, M. 1995, \textit{ApJ}, 451, 675

\bibitem[Elmegreen et al. (2009a)]{elmegreen09a} Elmegreen, B.G.,
Elmegreen, D.M., Fernandez, M.X., \& Lemonias, J.J. 2009a,
\textit{ApJ}, 692, 12

\bibitem[Elmegreen et al. (2009c)]{e09c} Elmegreen, B.G., Galliano, E.
\& Alloin, D. 2009, \textit{ApJ}, 703, 1297

\bibitem[D. Elmegreen (1980)]{e80} Elmegreen, D.M. 1980,
\textit{ApJS}, 43, 37

\bibitem[Elmegreen \& Elmegreen (1995)]{e95} Elmegreen, D.M., \&
Elmegreen, B.G. 1995 \textit{ApJ}, 445, 591

\bibitem[Elmegreen et al. (2009b)]{elmegreen09b} Elmegreen, D.M.,
Elmegreen, B.G., Marcus, M., et al. 2009b,  \textit{ApJ}, 701, 306

\bibitem[Elmegreen et al.(2011)]{elm11} Elmegreen, D.M., Elmegreen,
B.G., Yau, A. et al. 2011, ApJ, 737, 32

\bibitem[Gao \& Solomon(2004)]{gao04} Gao, Y. \& Solomon, P.M. 2004,
\textit{ApJ}, 609, 271

\bibitem[Genzel et al. (2011)]{genzel11} Genzel, R., Newman, S.,
Jones, T., et al. 2011, \textit{ApJ} , 733, 101

\bibitem[Gil de Paz et al. (2007)]{gil07} Gil de Paz, A. et al. 2007,
\textit{ApJS}, 173, 185

\bibitem[Gnedin \& Kravtsov (2011)]{gnedin11} Gnedin, N.Y., \&
Kravtsov, A.V. 2011, \textit{ApJ}, 728, 88

\bibitem[Grabelsky et al. (1987)]{grabelsky87} Grabelsky, D. A.,
Cohen, R. S., Bronfman, L., Thaddeus, P., \& May, J. 1987,
\textit{ApJ}, 315, 122

\bibitem[Gritschneder et al. (2009)]{grit09} Gritschneder, M., Naab,
T., Walch, S., Burkert, A., \& Heitsch, F. 2009, \textit{ApJ}, 694, L26

\bibitem[Heiner et al. (2011)]{heiner11} Heiner, J. S., Allen, R. J.,
van der Kruit, P. C. 2011, \textit{MNRAS}, 416, 2

\bibitem[Heitsch et al. (2008)]{heitsch08} Heitsch, F., Hartmann,
L.W., Burkert, A. 2008, \textit{ApJ}, 683, 786

\bibitem[Herrara et al. (2011)]{herrara11} Herrera, C. N., Boulanger,
F., \& Nesvadba, N. P. H. 2011, \textit{A\&A}, 534, 138

\bibitem[Hollis (1996)]{hollis96} Hollis, A. J. 1996, \textit{J.
British Astron. Assoc.}, 106, 354

\bibitem[Jog \& Solomon (1984)]{jog84} Jog, C.J., \& Solomon, P.M.
1984, \textit{ApJ}, 276, 114

\bibitem[Julian \& Toomre (1966)]{julian66} Julian, W.H., \& Toomre,
A. 1966, \textit{ApJ}, 146, 810

\bibitem[Kennicutt(1998)]{ken98} Kennicutt, R.C., Jr. 1998,
\textit{ApJ}, 498, 541

\bibitem[Kim \& Ostriker (2002)]{kim02a} Kim, W.-T., \& Ostriker, E.C.
2002, \textit{ApJ}, 570, 132

\bibitem[Kim et al. (2002)]{kim02} Kim, W.-T., Ostriker, E.C., \&
Stone, J.M. 2002, \textit{ApJ}, 581, 1080

\bibitem[Klein et al.(1983)]{klein83} Klein, R. I., Sandford, M. T.,
II, \& Whitaker, R. W. 1983, \textit{ApJL}, 271, 69

\bibitem[Konyves et al. (2007)]{konyves07} K\"onyves, V., Kiss, Cs.,
Mo\'or, A., Kiss, Z. T., \& T\'oth, L. V. 2007, \textit{A\&A}, 463,
1227

\bibitem[Lallement et al. (2003)]{lallement03} Lallement, R., Welsh,
B. Y., Vergely, J. L., Crifo, F., \& Sfeir, D. 2003, \textit{A\&A},
411, 447

\bibitem[Leroy et al.(2008)]{leroy08} Leroy, A.K., Walter, F., Brinks,
E. et al. 2008, \textit{AJ}, 136, 2782

\bibitem[Lin \& Shu (1964)]{lin64} Lin, C.C., \& Shu, F.H. 1964, ApJ,
140, 646

\bibitem[Lindblad et al. (1973)]{lind73} Lindblad, P. O., Grape, K.,
Sandqvist, A., \& Schober, J. 1973, \textit{A\&A}, 24, 309

\bibitem[Matsuda \& Nelson (1977)]{matsuda77} Matsuda, T. \& Nelson,
A.H. 1977, \textit{Nature}, 266, 607

\bibitem[Patel et al.(1995)]{patel95} Patel, N.A., Goldsmith, P.F.,
Snell, R.L., Hezel, T., \& Xie, T. 1995, \textit{ApJ}, 447, 721

\bibitem[Rafikov (2001)]{rafikov01} Rafikov, R.R. 2001,
\textit{MNRAS}, 323, 445

\bibitem[Reach et al.(2009)]{reach09} Reach, W.T., et al. 2009,
\textit{ApJ}, 690, 683

\bibitem[Roberts(1969)]{roberts69} Roberts, W. W. 1969, \textit{ApJ},
158, 123

\bibitem[Romeo \& Wiegert (2011)]{romeo11} Romeo, A.B., \& Wiegert, J.
2011, \textit{MNRAS}, 416, 1191

\bibitem[Smith(2010)]{smith10} Smith, N. 2010, \textit{MNRAS}, 406,
952

\bibitem[Sugitani et al.(1989)]{sugitani89} Sugitani, K., Fukui, Y.,
Mizuni, A., \& Ohashi, N. 1989, \textit{ApJL}, 342, 87

\bibitem[Sugitani et al.(1995)]{sugitani95} Sugitani, K., Tamura, M.,
Ogura, K. 1995, \textit{ApJL}, 455, 39

\bibitem[Tacconi et al. (2010)]{tacc10} Tacconi, L., et al. 2010,
\textit{Nature}, 463, 781

\bibitem[Toomre (1969)]{toomre69} Toomre, A. 1969, ApJ, 158, 899

\bibitem[Toomre(1981)]{toomre81} Toomre, A. 1981, in: S.M. Fall (ed.),
\textit{The structure and evolution of normal galaxies}, (Cambridge:
Cambridge University Press), p.\,111.

\bibitem[Wu et al.(2005)]{wu05} Wu, J., Evans, N.J., II., Gao, Y.,
et al. 2005, \textit{ApJ}, 635, L173

\bibitem[Zavagno et al.(2006)]{zavagno06} Zavagno, A., Deharveng, L.,
Comer\'on, F., et al. 2006, \textit{A\&A}, 446, 171

\end{thebibliography}
\end{document}